%% file: rayleigh_interface6.tex
\documentclass[11pt]{article}

\usepackage[top=1.0in,right=1.0in,left=1.0in,bottom=1.75in]{geometry}
\usepackage{amssymb,amsbsy}
\usepackage{amsmath}
\usepackage{amsthm}
\usepackage{graphics,graphicx}
\usepackage[T1]{fontenc}
\usepackage{color}
\usepackage{subfigure}
\usepackage[affil-it,auth-sc]{authblk}
\usepackage{stmaryrd}
\usepackage{amsfonts}
\usepackage{amssymb}

\usepackage[]{jabbrv}

\usepackage[numbers,sort&compress]{natbib}

\input{definitions}

\graphicspath{{./}{./figures/}}

\title{Rotational inertia interface in \\ a dynamic lattice of flexural beams}

\author[1]{A. Piccolroaz\footnote{Corresponding author: e-mail: roaz@ing.unitn.it; phone: +39\,0461\,282583.}}
\author[2]{A.B. Movchan}
\author[1]{L. Cabras}

\affil[1]{Dipartimento di Ingegneria Meccanica e Strutturale, Universit\`a di Trento, Italy}
\affil[2]{Department of Mathematical Sciences, University of Liverpool, U.K.}

\date{}

\begin{document}

\maketitle

\begin{abstract}
\noindent
The paper presents a novel analysis of a transmission problem for a network of flexural beams incorporating conventional Euler-Bernoulli beams as well as Rayleigh beams with the enhanced rotational inertia.
Although, in the low-frequency regime, these beams have a similar dynamic response, we have demonstrated novel features which occur in the transmission at higher frequencies across the layer of the Rayleigh beams.

\end{abstract}





\section{Introduction}
\label{sec01}

The paper addresses modelling of a structured interface built of a network of flexural beams with an enhanced rotational inertia. We refer to these as Rayleigh beams, and in the static limit they are indistinguishable from the classical Euler-Bernoulli beams, but at higher frequency regimes show remarkable dispersion properties.
Problems of homogenisation for differential operators with rapidly oscillating coefficients are of high interest in the applied mathematics and physics community (see, for example, Bensoussan {\em et al.}~\cite{BLP1978}, Sanchez-Palencia \cite{SP1980}, Jikov {\em et al.}~\cite{Jikov1994}). In particular, modelling for structures and composites, which include networks of elastic rods and beams, was considered by Panasenko \cite{Panasenko2005}.

Exciting results concerning wave propagation in two-dimensional periodic lattices, and in particular filtering, were obtained by Srikantha Phani {\em et al.}~\cite{Fleck2006}, and effects of dynamic anisotropy were highlighted by Antonakakis {\em et al.}~\cite{Craster_2014}.

Structured interfaces in dynamics and statics, in the context of elastic systems, trapped waveforms, and resonances, were considered by Bigoni and Movchan \cite{Bigoni_2002}.

The modelling of Floquet-Bloch waves for lattice structures consisting of flexural beams attracted the attention of many authors in the past in conjunction with a wide range of applications in physics and mechanics. In particular,  the analytical approach by Brun {\em et al.}~\cite{Brun_2013} has proved to be efficient for the analysis of propagation of damage, which was motivated by the earlier work on fracture in lattices by  Slepyan \cite{Slepyan_2002} and Slepyan and Ryvkin \cite{Slepyan_Ryvkin_2010}. For one-dimensional periodic structures, consisting of elastic beams, closed forms analytical solutions were derived and analysed by
Heckl \cite{Heckl_2002}.

The networks of Rayleigh beams and the effects of rotational inertia on the dispersion of Floquet waves have been studied by Piccolroaz and Movchan \cite{PM_2014}, and the novel studies of dynamic anisotropy for two-dimensional networks incorporating both the Rayleigh and Euler-Bernoulli beams were included in the recent paper by Piccolroaz, Movchan and Cabras \cite{PMC_2016}.

\begin{figure}[!htb]
\centering
\includegraphics[width=160mm]{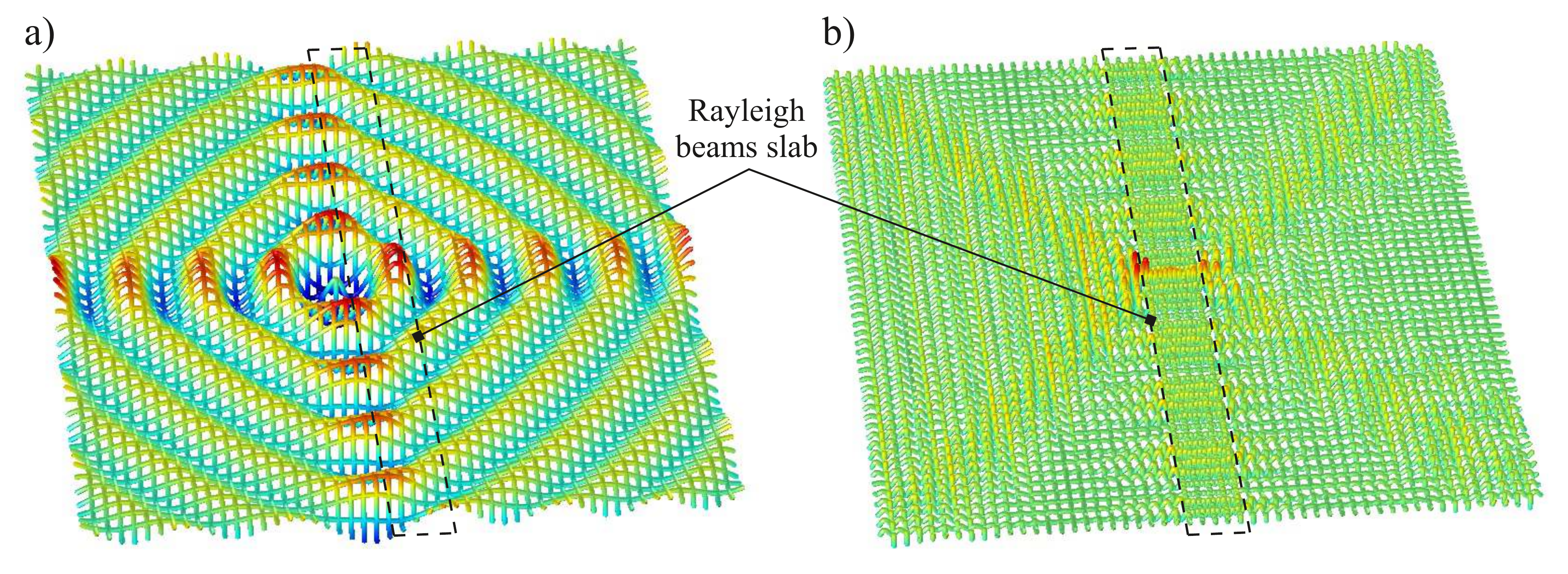}
\caption{\footnotesize Wave propagating in a Euler-Bernoulli beam network and interacting with a structured interface, made of Rayleigh beams. a) Low frequency regime $\omega=0.06\!\cdot\!\pi$ rad/s. b) High frequency regime $\omega=2.54\!\cdot\!\pi$ rad/s.}
\label{figINTRO}
\end{figure}

In Fig.~\ref{figINTRO}, we give an illustrative computation for a wave propagating in a Euler-Bernoulli beam network and interacting with a structured interface, made of Rayleigh beams. At low frequencies, the interface is invisible, as the Rayleigh beams respond in a way similar to the classical Euler-Bernoulli beams, as shown in Fig.~\ref{figINTRO}a. However, negative refraction and focussing across the interface are observed at higher frequencies, as shown in Fig.~\ref{figINTRO}b.

The structure of the paper is as follows. In Section \ref{sec02} we discuss the governing equations for the mathematical model, as well as the geometry of the structured interface.
The auxiliary important material concerning with the dispersion of Floquet-Bloch waves in periodic structures is discussed in Section  \ref{sec03}.
Finally, the main results of the computations analysing the dynamic response of the structured interface, which possesses an enhanced rotational inertia, are included in Section \ref{sec04}.
We also note that the numerical computations, produced in COMSOL Multiphysics, required absorbing non-reflective boundary conditions. The Appendix includes the technical details concerning with the derivation and implementation of these boundary conditions.

\section{Governing equations and geometry of the interface}
\label{sec02}

\begin{figure}[!htb]
\centering
\includegraphics[width=140mm]{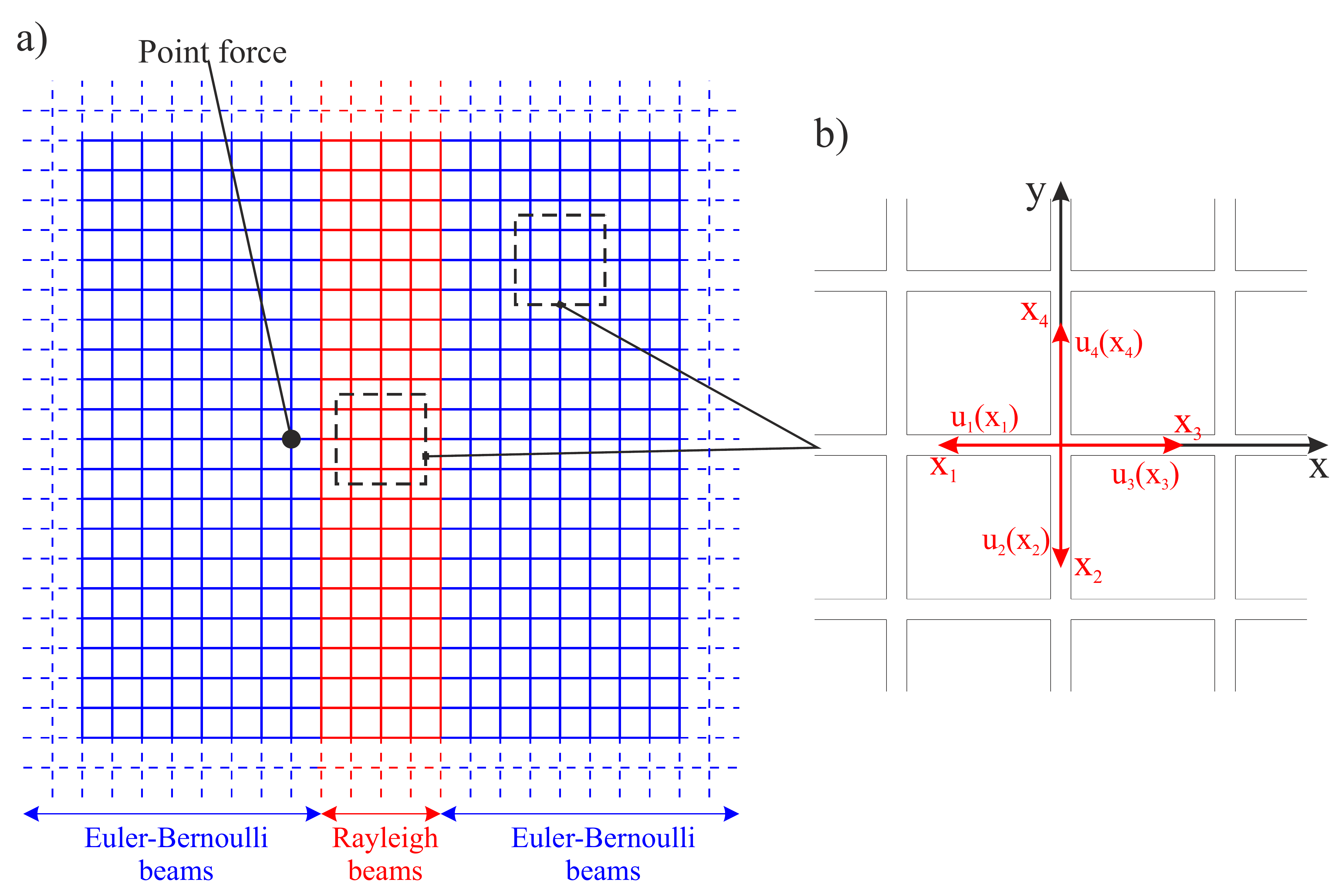}
\caption{\footnotesize a) An infinite rectangular periodic network of Euler-Bernoulli beams with a structured interface made of Rayleigh beams. b) A zoom of the dashed squares showing the local coordinates for each beam element.}
\label{fig01}
\end{figure}

We consider an infinite rectangular periodic network of Euler-Bernoulli beams with a structured interface made of Rayleigh beams, as shown in Fig.~\ref{fig01}, and assume that the beams can deflect in the out-of-plane direction.

The governing equation for time-harmonic flexural waves in a Rayleigh beam is
\begin{equation}
\label{eq:gov}
EI u''''(x) + \rho I \omega^2 u''(x)  - \rho A \omega^2 u(x) = 0,
\end{equation}
where $E$ is the Young modulus,  $\rho$ the mass density, $A$ the area of the cross-section, and $I$ the area moment of inertia of the cross-section. 

The internal bending moment $M$ and the internal shear force $V$ are given by
\begin{equation}
M(x) = -EI u''(x), \quad V(x) = -EI u'''(x)  - \rho I \omega^2 u'(x),
\end{equation}
respectively.

Assuming a flexural wave satisfying (\ref{eq:gov})  in the form
\begin{equation}
u(x) = C e^{i \kappa x},
\end{equation}
the characteristic roots $k$ are written in the form
\begin{equation}
\kappa_{1,2,3,4} = \pm \sqrt{\frac{\rho I\omega^2 \pm \omega\sqrt{\rho^2I^2\omega^2+4EI\rho A}}{2 EI}}.
\end{equation}
When there are no multiple characteristic roots, the flexural wave is then given by the linear combination
\begin{equation}
u(x) = \sum_{q=1}^{4} C_q e^{i \kappa_q x},
\end{equation}

This representation is used for waves inside the Rayleigh beam slab of finite thickness, as shown in Fig.~\ref{fig01}. We also refer to this region as the ``structured interface''. Outside the structured interface, the square network of beams consists of classical Euler-Bernoulli beams, described by the equation (\ref{eq:gov}), where the rotational inertia term is absent, i.e.
\begin{equation}
\label{eq:govv}
EI u''''(x)  - \rho A \omega^2 u(x) = 0 \ \ \ \mbox{for an Euler-Bernoulli beam}.
\end{equation}
The classical balance conditions for forces and moments are set at the nodal points of the lattice, whereas the flexural displacements and their first-order derivatives are continuous throughout the lattice.

The problem considered here is inhomogeneous, with a time-harmonic point force being applied in the close proximity of the structured interface, to a nodal point inside the Euler-Bernoulli flexural lattice. The applied force acts in the out-of-plane direction, perpendicular to the $(x, y)$--plane.

\section{Dispersion of flexural waves in periodic networks}
\label{sec03}

Here we outline the results of the earlier work by Piccolroaz, Movchan and Cabras \cite{PMC_2016}, which dealt with the dispersion properties of Floquet-Bloch waves in periodic lattices of flexural beams.

\begin{figure}[!htb]
\centering
\includegraphics[width=120mm]{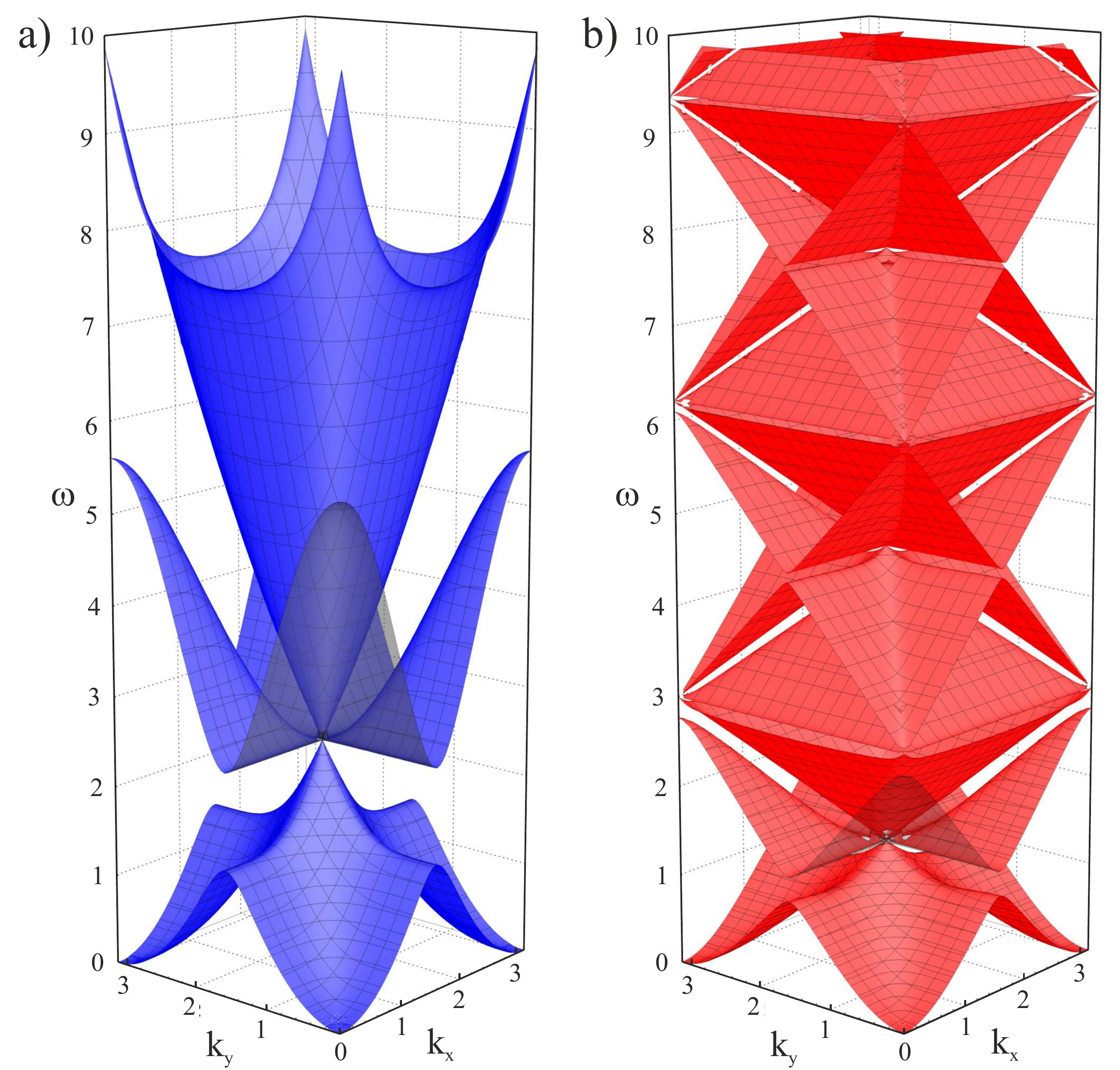}
\caption{\footnotesize a) Dispersion surface for the Euler-Bernoulli  beam structure. b) Dispersion surface for the Rayleigh beam structure.
We note that the effect of rotational inertia in (b) is significant: the first three dispersion surfaces shown here occur at much lower frequencies compared to the corresponding surfaces for the Euler-Bernoulli's beams as in part (a). Dirac cones are clearly visible in both cases. However, the dispersion profiles near the Dirac cone vertices are different for the Rayleigh beam structure and for the Euler-Bernoulli beam structure. The following parameters have been used: $E=1$, $\rho=1$, $A=1$, $I=1$, $L=2$.}
\label{fig3d}
\end{figure}

\begin{figure}[!htb]
\centering
\includegraphics[width=160mm]{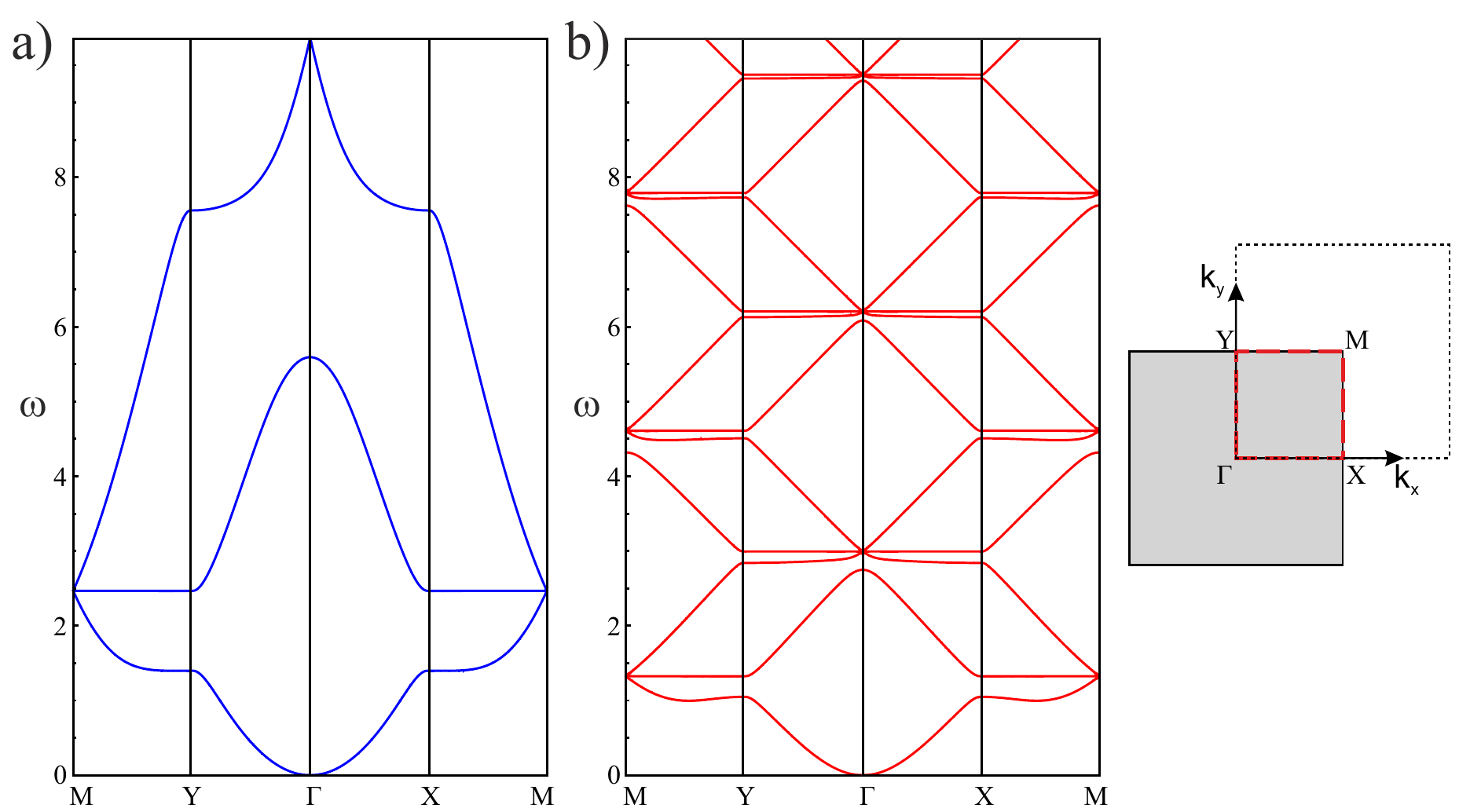}
\caption{\footnotesize The cross-sectional dispersion diagrams on the boundary of the Brillouin zone, for Floquet-Bloch waves in the networks of the Euler-Bernoulli beams (part (a) of the figure) and the Rayleigh beams (part (b) of the figure). The inset on the right shows the contour $\Gamma XMY$ within the first Brillouin zone in the elementary cell of the reciprocal lattice; the dotted square corresponds to the computational window chosen to draw the dispersion surfaces in Fig.\ref{fig3d}.}
\label{figband2}
\end{figure}

The Figures \ref{fig3d} and \ref{figband2} present the dispersion surface and the cross-sectional plots for two types of lattices: the Euler-Bernoulli flexural lattice (part (a) of the figure) and the Rayleigh lattice (part (b) of the figure).
As seen in these figures, the systems are both orthotropic and exhibit the same response in the low frequency range. However, at higher frequencies their response becomes very different. In particular, we pay a special attention to the so-called hyperbolic dynamic regimes, which correspond to the neighbourhoods of the saddle points on the dispersion surface. These phenomena are discussed in detail in Piccolroaz, Movchan and Cabras \cite{PMC_2016}, and here we note that two intersecting Gaussian-type beams are identified in the wave pattern, which corresponds to a hyperbolic mode.

Compared to the earlier paper \cite{PMC_2016}, the diagrams in Figs.~\ref{fig3d} and \ref{figband2} have been updated to the higher frequencies for the case of the Rayleigh beams. In particular, the cross-sectional plot in Fig.~\ref{figband2}b shows the pattern of Dirac cones extending to the higher frequency range. The interval of frequencies range in the immediate neighbourhood of the Dirac cones corresponds to the strong dynamic anisotropy, with the two orthogonal preferential directions. We also note the presence of standing waves, at the point $\Gamma$, corresponding to ${\bf k} = {\bf 0},$ in close proximity of the Dirac cones. With the increase of the frequency range, the standing wave becomes closer to  the frequency of the corresponding Dirac cone.

\clearpage

\section{Computational model of the dynamic flexural interface}
\label{sec04}

This section includes the results of the computational model, which provides the novel highlights in the dynamic response of the interface between the networks of the Euler-Bernoulli and of the Rayleigh beams.
The excitation is carried out by a time-harmonic out-of-plane point force placed at a nodal point in the network of the Euler-Bernoulli beams, adjacent to the boundary of the structured interface.

In order to simulate an infinite lattice with a finite-size computational window, a special care must be taken in defining the conditions at the exterior boundary. Simple supported or clamped boundary conditions would produce unreliable results, because the reflected waves interact with the incident waves. We have developed special absorbing boundary conditions to be imposed at the exterior contour of the computational window. The derivation of such boundary conditions for both Euler-Bernoulli and Rayleigh beams is included in the technical appendix.

The computations presented in this section have been performed with the same values of the parameters as those used for the dispersion surfaces shown in Fig.~\ref{fig3d}.

\subsection{Low-frequency response}

\begin{figure}[!htb]
\centering
\includegraphics[width=120mm]{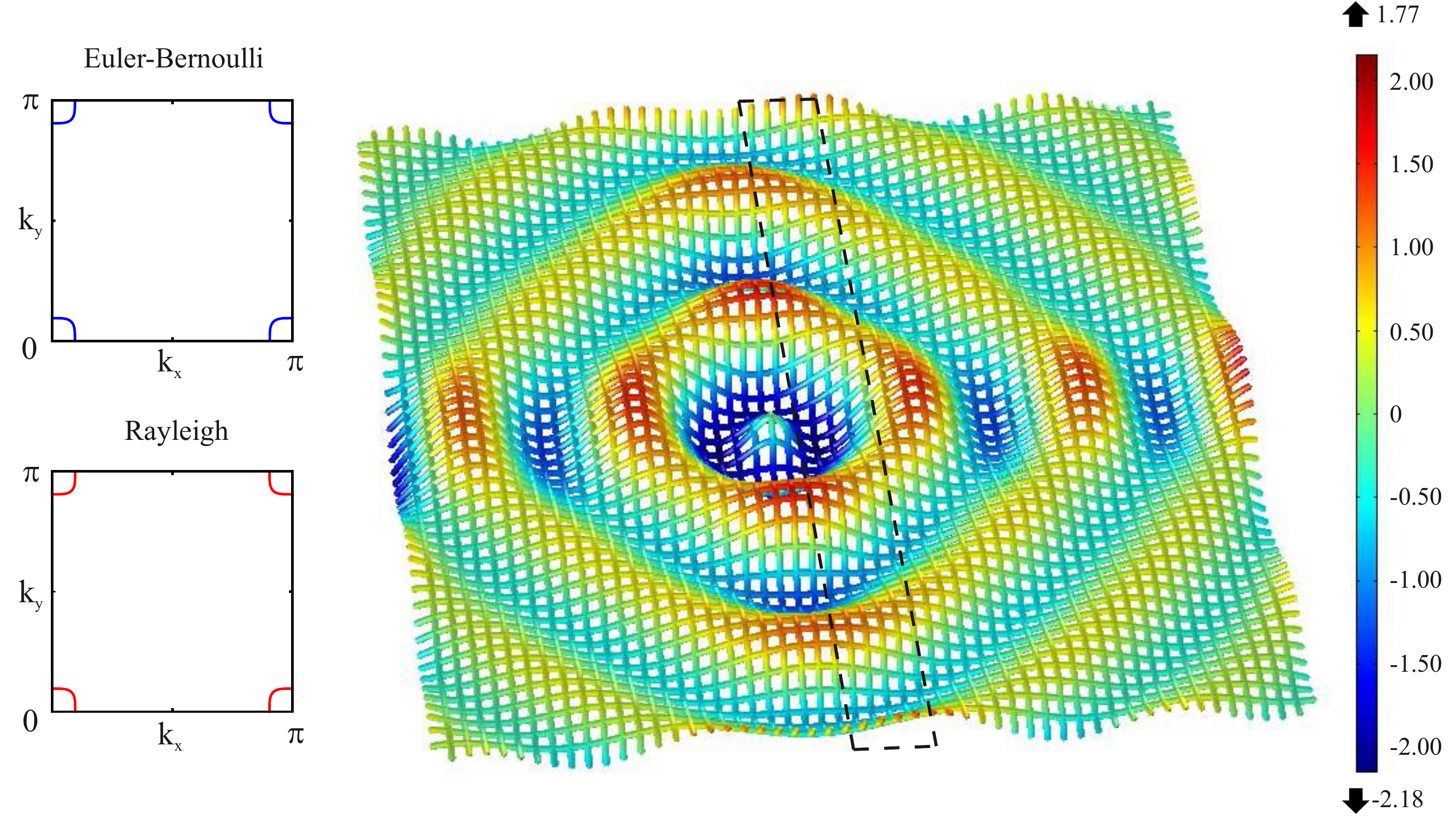}
\caption{\footnotesize
Long-wave illustration. In this regime the Euler-Bernoulli and Rayleigh beams are hardly distinguishable and the wave propagates as in the homogeneous lattice.}
\label{Rayleigh0001}
\end{figure}

In Fig.~\ref{Rayleigh0001}, we see the wave pattern exerted by a point force adjacent to the boundary of the structured interface, at the angular frequency of $0.02\!\cdot\!\pi$ rad/s. We observe a typical behaviour for a homogeneous orthotropic structure, and the presence of the interface is not visible, as both the Rayleigh and the Euler-Bernoulli beams exhibit similar dynamic response.

The inset of Fig.~\ref{Rayleigh0001} also includes two slowness contours obtained for the networks of the Euler-Bernoulli and Rayleigh beams at the angular frequency of $0.02\!\cdot\!\pi$ rad/s. These slowness contours are indistinguishable, as expected, since the low-frequency response of the square networks of the Rayleigh and Euler-Bernoulli beams is similar. The wave pattern shown in the Figure is a typical one for orthotropic flexural networks.

\subsection{Uni-directional reflection from the structured interface}

\begin{figure}[!htb]
\centering
\includegraphics[width=120mm]{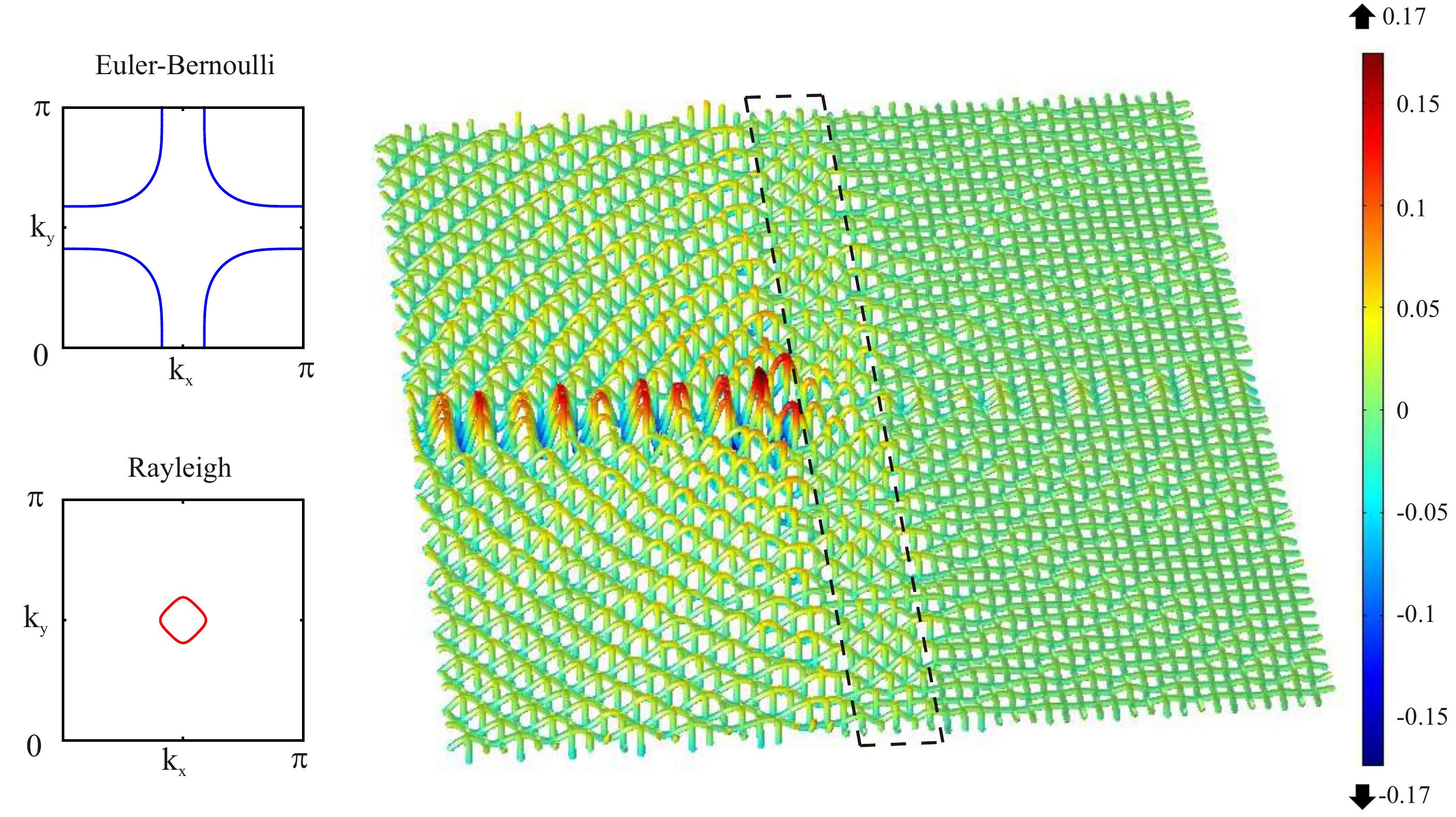}
\caption{\footnotesize
Uni-directional reflection $\omega=0.36\!\cdot\!\pi$ rad/s.}
\label{Rayleigh0018}
\end{figure}

With the angular frequency increase up to $0.36\!\cdot\!\pi$ rad/s, we observe a new feature, linked to the so-called hyperbolic dynamic regime. The network of the Euler-Bernoulli beams has a typical cross-shaped wave pattern aligned with the coordinate axes, whereas the network of the Rayleigh beams has the cross-shaped wave pattern rotated through the angle of $45^\circ$. In the transmission problem for the structured interface with the enhanced rotational inertia, this results in a uni-directional reflection, as shown in Fig.~\ref{Rayleigh0018}.

This unusual dynamic response of the structured interface is also demonstrated through the slowness contours, included as the inset of Fig.~\ref{Rayleigh0018}. These slowness contours appear to be very different for the networks of the Rayleigh and of the Euler-Bernoulli beams, in particular in the choice of the preferential directions.

\subsection{Negative refraction and focussing across the structured interface with the enhanced rotational inertia}

The effects of strong dynamic anisotropy can also be visualised through a negative refraction across the structured interface. For electromagnetic and acoustic waves such phenomena are well-known (see, for example, \cite{JBP, PSS, CG}). However, for flexural waves at the interfaces between the Rayleigh and Euler-Bernoulli beams this has never been addressed.

\begin{figure}[!htb]
\centering
\includegraphics[width=160mm]{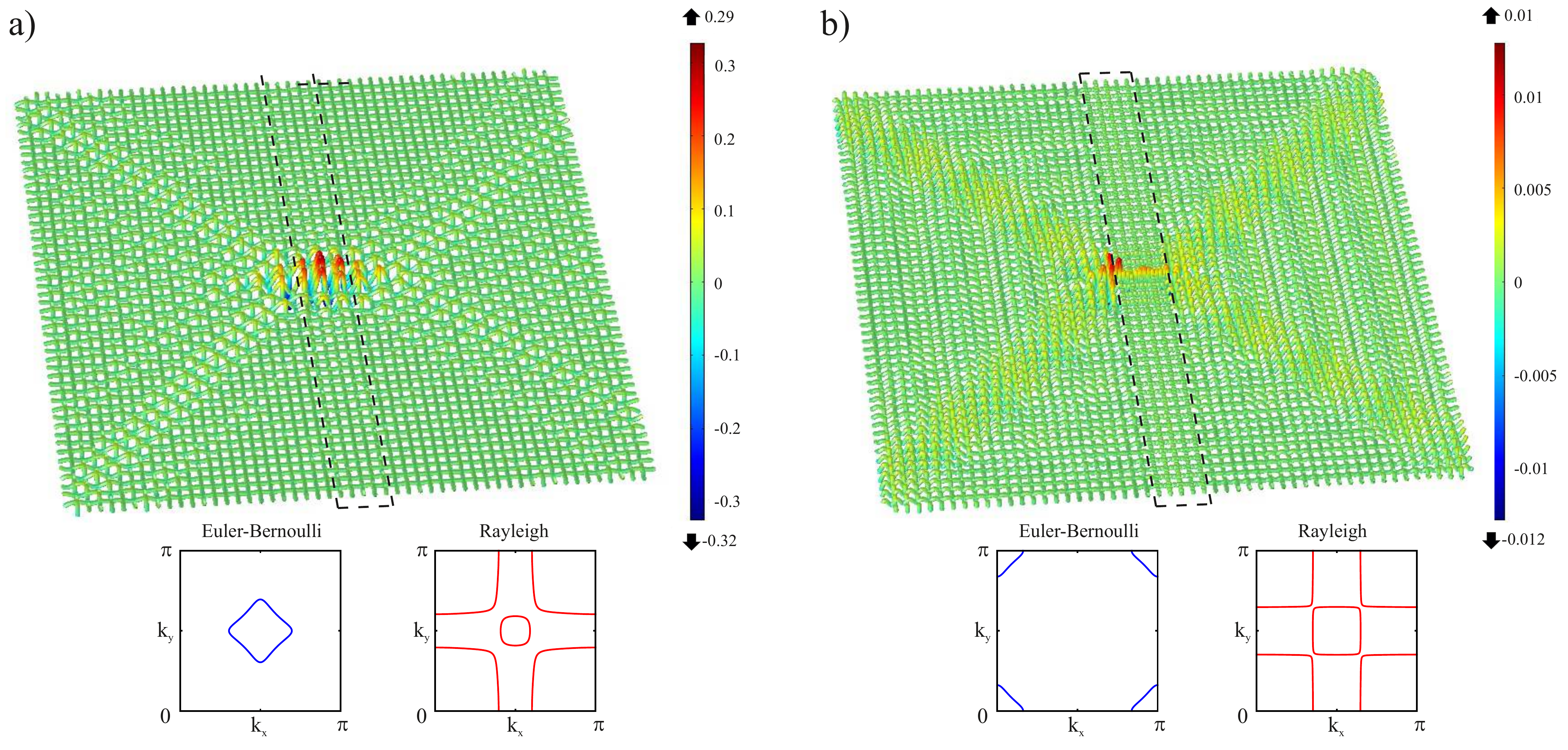}
\caption{\footnotesize
a) Focussing $\omega=0.50\!\cdot\!\pi$ rad/s. b) Focussing $\omega=2.62\!\cdot\!\pi$ rad/s.}
\label{Rayleighfocusing}
\end{figure}

The figures \ref{Rayleighfocusing}a and \ref{Rayleighfocusing}b were obtained at the angular frequencies of $0.50\!\cdot\!\pi$ rad/s and $2.62\!\cdot\!\pi$ rad/s, respectively. They both exhibit the phenomena of negative refraction and focussing across the structured interface of flexural Rayleigh beams.
At the lower angular frequency of $0.50\!\cdot\!\pi$ rad/s the wave pattern is highly localised within the interface (see Fig.~\ref{Rayleighfocusing}a), whereas at the higher angular frequency of $2.62\!\cdot\!\pi$ rad/s the focussing is also accompanied by symmetric pattern, as shown in Fig.~\ref{Rayleighfocusing}b corresponding to a hyperbolic dynamic regime (a cross-like wave pattern rotated through $45^\circ$ relative to the coordinate axes).

The iso-frequency maps, shown in the inset of the Fig.~\ref{Rayleighfocusing}, show different slowness contours for the networks of the Rayleigh and Euler-Bernoulli beams, which also results in the different orientation of the preferential directions for flexural waves at the given angular frequencies of $0.50\!\cdot\!\pi$ rad/s and $2.62\!\cdot\!\pi$ rad/s.


\subsection{Waveguide modes within the structured interface }

\begin{figure}[!htb]
\centering
\includegraphics[width=160mm]{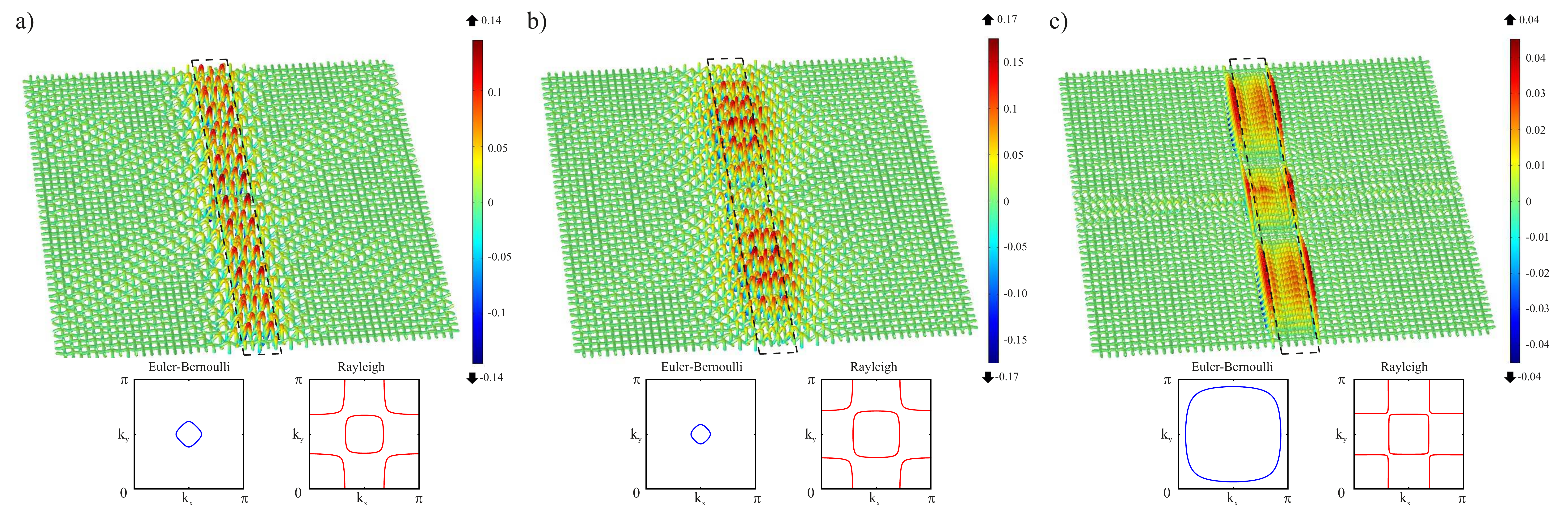}
\caption{\footnotesize
a) Waveguide mode $\omega=0.58\!\cdot\!\pi$ rad/s. b) Waveguide mode $\omega=0.62\!\cdot\!\pi$ rad/s. c) Waveguide mode $\omega=2.28\!\cdot\!\pi$ rad/s.}
\label{Rayleighwaveguide}
\end{figure}

Three figures \ref{Rayleighwaveguide}a, \ref{Rayleighwaveguide}b and \ref{Rayleighwaveguide}c, constructed for the excitation angular frequencies of
$0.58\!\cdot\!\pi$ rad/s, $0.62\!\cdot\!\pi$ rad/s and $2.28\!\cdot\!\pi$ rad/s, respectively, show the wave patterns entirely localised within the interface consisting of the Rayleigh beams. Although the vibration was induced by the point force adjacent and outside of the structured interface, the wave pattern is clearly showing the wave propagating in the vertical direction and localised within the structured interface. The intensity of the localised waveform and the modulation pattern depend on the frequency of the excitation, as demonstrated here.

The slowness contour diagrams included in the insets for these figures, indicate that there are no slowness contours enclosing the origin $k_x=k_y=0$ at the chosen angular frequencies for the flexural waves in the Euler-Bernoulli lattice. Preferential directional along the coordinate axes are clearly shown in the slowness contours for the networks of the Rayleigh beams. The result of an interaction between such networks is a highly localised waveguide mode within the structured interface; the shape of the modulating function depends on the frequency, as expected.

\newpage

\subsection{Localisation}

\begin{figure}[!htb]
\centering
\includegraphics[width=160mm]{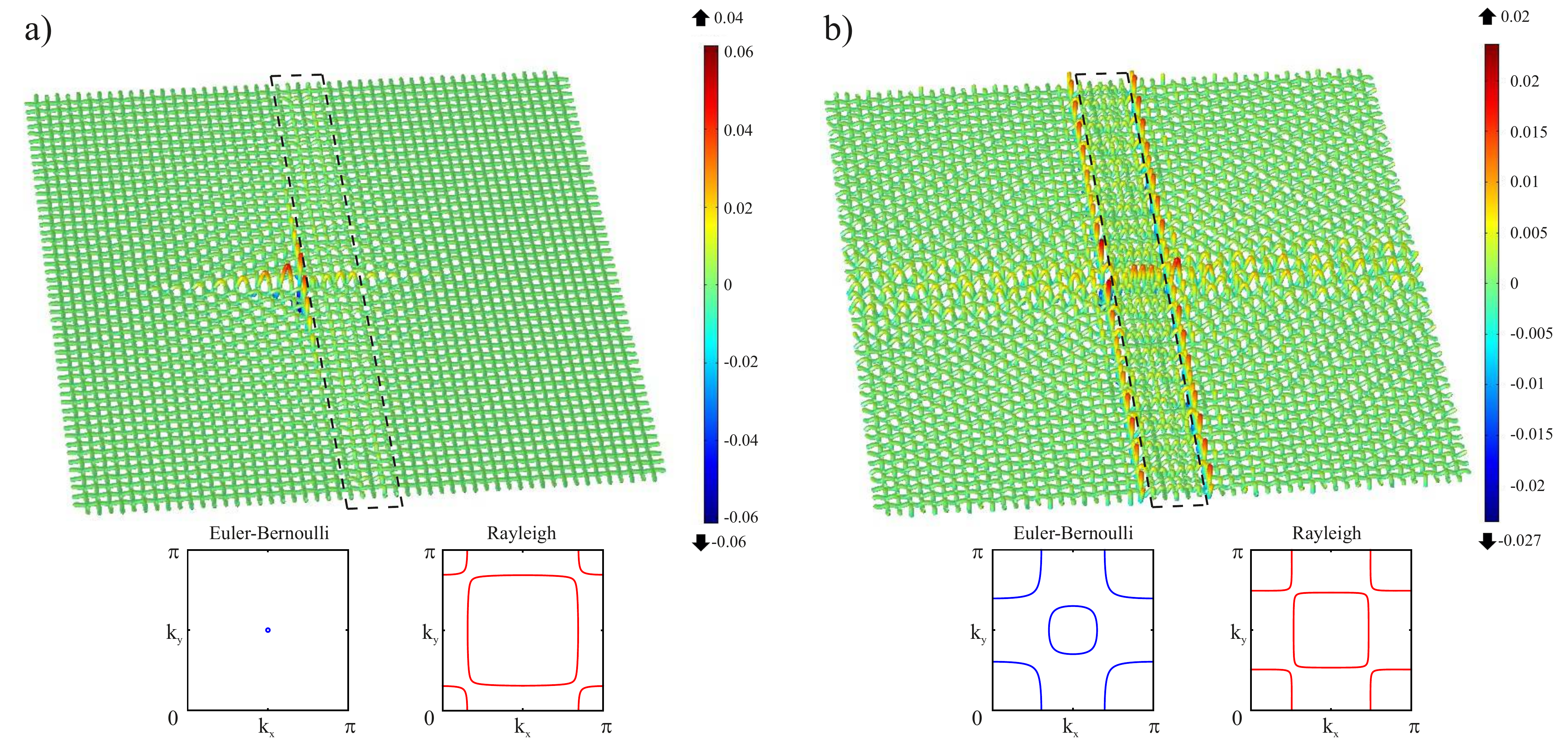}
\caption{\footnotesize
a) Localisation $\omega=0.76\!\cdot\!\pi$ rad/s. b) Interfacial localisation $\omega=1.20\!\cdot\!\pi$ rad/s.}
\label{Rayleighlocalisation}
\end{figure}

Localisation around the interface with the enhanced rotational inertia is a very interesting phenomenon, which may occur at different frequencies. Here we provide an illustration based on two examples constructed for angular frequencies of $0.76\!\cdot\!\pi$ rad/s and $1.20\!\cdot\!\pi$ rad/s. In Fig.~\ref{Rayleighlocalisation}a, the localisation occurs near the point of excitation, and the wave pattern is confined within a neighbourhood of that point. On the contrary, the localisation, shown at a higher frequency in Fig.~\ref{Rayleighlocalisation}b, shows a wave pattern localised as an interface wave, along the boundary separating the networks of Euler-Bernoulli and of the Rayleigh beams.

The two cases, referred to as ``localisation'' and shown in Fig.~\ref{Rayleighlocalisation}, are very different, as seen from the corresponding slowness contours. The part (a) of the Figure corresponds to the ``Dirac cone'' regime in the network of the Euler-Bernoulli beams, with no slowness contours enclosing the origin. In this case the wave form is localised in the neighbourhood of the point where the point force has been applied, and within the structured interface. In the part (b), by ``localisation'' we mean an ``edge waveform'' which exists along the boundary separating the region occupied by the Euler-Bernoulli beams and the structured interface; the overall wave is not localised in the transverse direction across the interface.

\subsection{Resonance transmission}

Finally, we show an example of the resonance transmission in the high frequency regime, when the localisation of the wave inside the structured interface is also accompanied by the low reflection, and the relatively high transmission across the interface.

\begin{figure}[!htb]
\centering
\includegraphics[width=120mm]{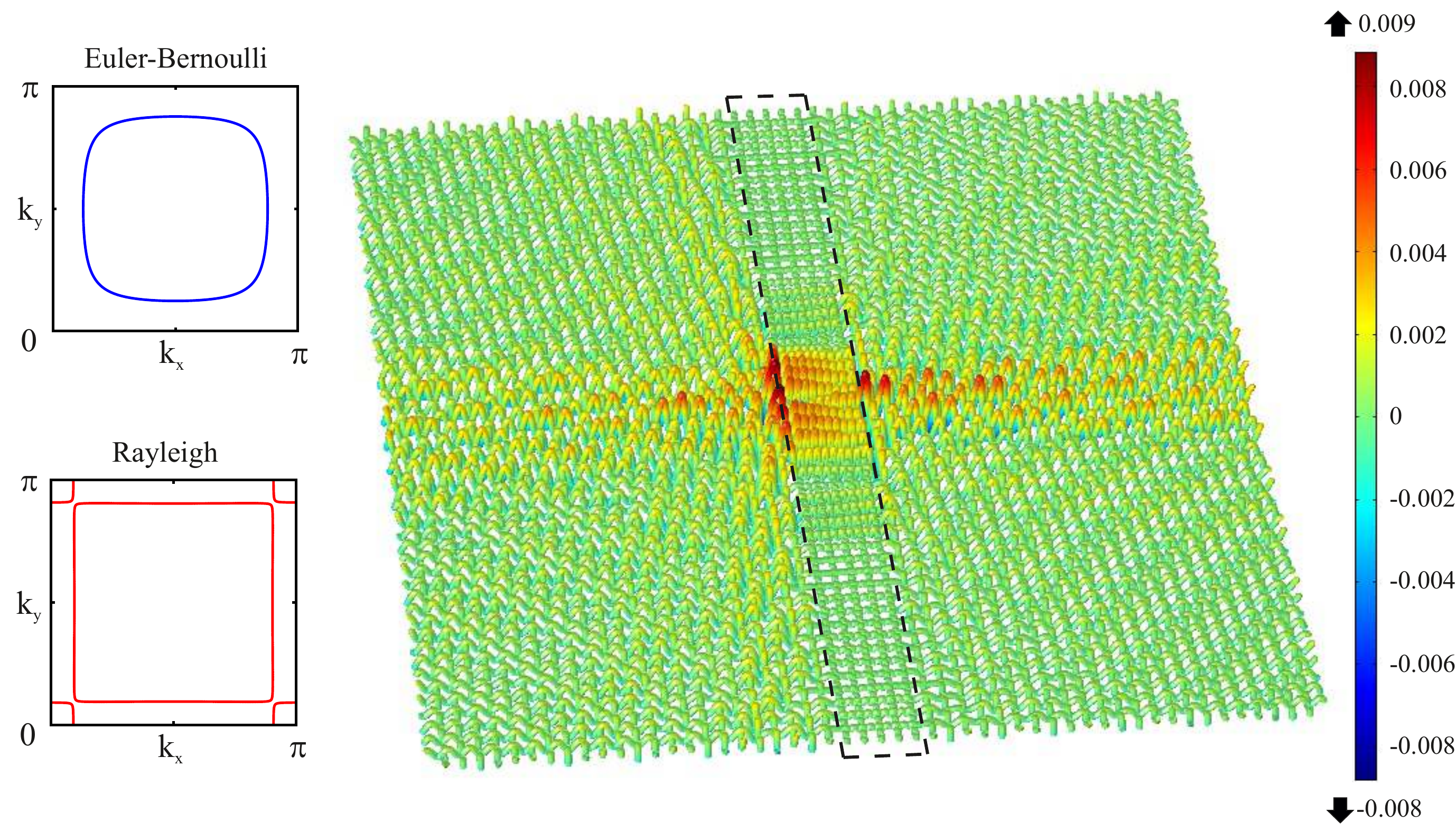}
\caption{\footnotesize
Resonance transmission $\omega=2.06\!\cdot\!\pi$ rad/s.}
\label{Rayleighresonance}
\end{figure}

Fig.~\ref{Rayleighresonance} shows the case when the localisation within the interface is  confined vertically,
and the wave pattern shows a high transmission rate and a uni-directional preference along the horizontal axis.

The slowness contours attached in the inset for this figure, indicate that no slowness contours enclose the origin in the case of the network of the Euler-Bernoulli beams. On the contrary, the preferential directions along the coordinate axes are clearly shown in the case of the flexural network of the Rayleigh beams.

\section{Concluding remarks}
\label{sec05}

The theory and computations presented in this paper have demonstrated the importance of the rotational inertia in controlling of flexural waves in structured media.

From the point of view of a quasi-static analysis, the prediction is counter-intuitive, since the Rayleigh and Euler-Bernoulli beams provide similar dynamic response in the low-frequency regimes. This makes the problem even more interesting, as one discovers how different these structures are at higher frequencies, especially in the cases of strong dynamic anisotropy.

The structured interface built of the Rayleigh beams is a very fascinating object, which has been demonstrated to have the properties of a waveguide, a flat lens, or a mirror for different frequency regimes of the elastic flexural waves.

Further possible directions, where the present  study can be extended is in the construction of the shields (or approximate cloaks) made of the networks of the Rayleigh beams.  Also lensing and mirroring of flexural waves in lattice, which include a combination of the Euler-Bernoulli and of the Rayleigh beams, becomes a straightforward task, which can build upon the results of the present study.

\vspace{6mm}
{\bf Acknowledgements}. AP would like to acknowledge financial support from the
European Union's Seventh Framework Programme FP7/2007-2013/ under REA grant
agreement number PCIG13-GA-2013-618375-MeMic. 
AM has visited the University of Trento in 2016 with the support from the
European Union's Grant ERC-2013-ADG-340561-INSTABILITIES, which is gratefully acknowledged.
AM also acknowledges support from the UK EPSRC Program Grant  EP/L024926/1.
LC acknowledges financial support from the University of Trento, within the research project 2014 entitled ``3D printed metallic foams for biomedical applications: understanding and improving their mechanical behavior''.

\clearpage
\appendix
\renewcommand{\theequation}{\thesection.\arabic{equation}}

\begin{center}
\bf \Large Appendix
\end{center}

\section{Non-reflective boundary conditions for dynamic flexural beams}
\setcounter{equation}{0}

In this paper, we modelled a lattice built of a network of flexural beams and we analysed its dynamic response. To do that, we introduced on the boundaries of the vibrating domain approximated wave-absorbing boundary conditions, that consists of minimizing the magnitude of reflected and evanescent waves due to the boundary effects. In the case of a vibrating string, it is possible to have complete absorption, realizing the virtual continuation of the string. However, for a flexural beam, we suppress reflected waves due to the corresponding incident waves but there still exists evanescent waves in the model.

\subsection{Euler-Bernoulli beam}

We start with the absorbing boundary condition for the Euler-Bernoulli beams. The governing equation for time-harmonic flexural waves in a thin flexible beam is:
\begin{equation}
\label{eq:gov1}
u''''(x)-\beta^2 u(x) = 0,
\end{equation}
where $\beta=\sqrt{\frac{\rho A}{E I}\omega^2}$. The general solution of the equation (\ref{eq:gov1}) is:
\begin{equation}
\label{eq:gov3}
u(x)=C_1e^{i\sqrt{\beta} x}+C_2e^{-i\sqrt{\beta} x}+C_3e^{\sqrt{\beta} x}+C_4e^{-\sqrt{\beta} x},
\end{equation}
where $C_1e^{i\sqrt{\beta} x}$ represents the travelling wave from left to right, $C_2e^{-i\sqrt{\beta} x}$ the travelling wave from right to left, $C_3e^{\sqrt{\beta} x}$ the evanescent wave decaying at $-\infty$ and $C_4e^{-\sqrt{\beta} x}$ the evanescent wave decaying at $+\infty$. For the reflection problem in the semi-infinite beam, in the domain $(-\infty,0)$, the coefficient $C_1$, $C_2$, $C_3$ and $C_4$ represents the amplitudes of the incident, reflected, and of the two evanescent waves respectively.

We consider the case where the coefficient responsible for the incident wave is $C_1^*=1$ and the one responsible for the second evanescent wave is $C_4^*=0$, and we determine two boundary conditions in $x=0$ in order to evaluate the two remaining coefficients $C_2^*$ and $C_3^*$:
\begin{equation}
\label{eq:gov4}
u''(0)=a_1 i \beta u(0) \quad \textrm{and} \quad u'''(0)=a_2 i \beta u'(0).
\end{equation}
The aim is to determine the parameters $a_2$ and $a_1$ such that the coefficients $C_2^*$ and $C_3^*$, responsible for the reflected and evanescent waves respectively, are small and independent of the pulsation. 

\begin{figure}[!htb]
\centering
\includegraphics[width=160mm]{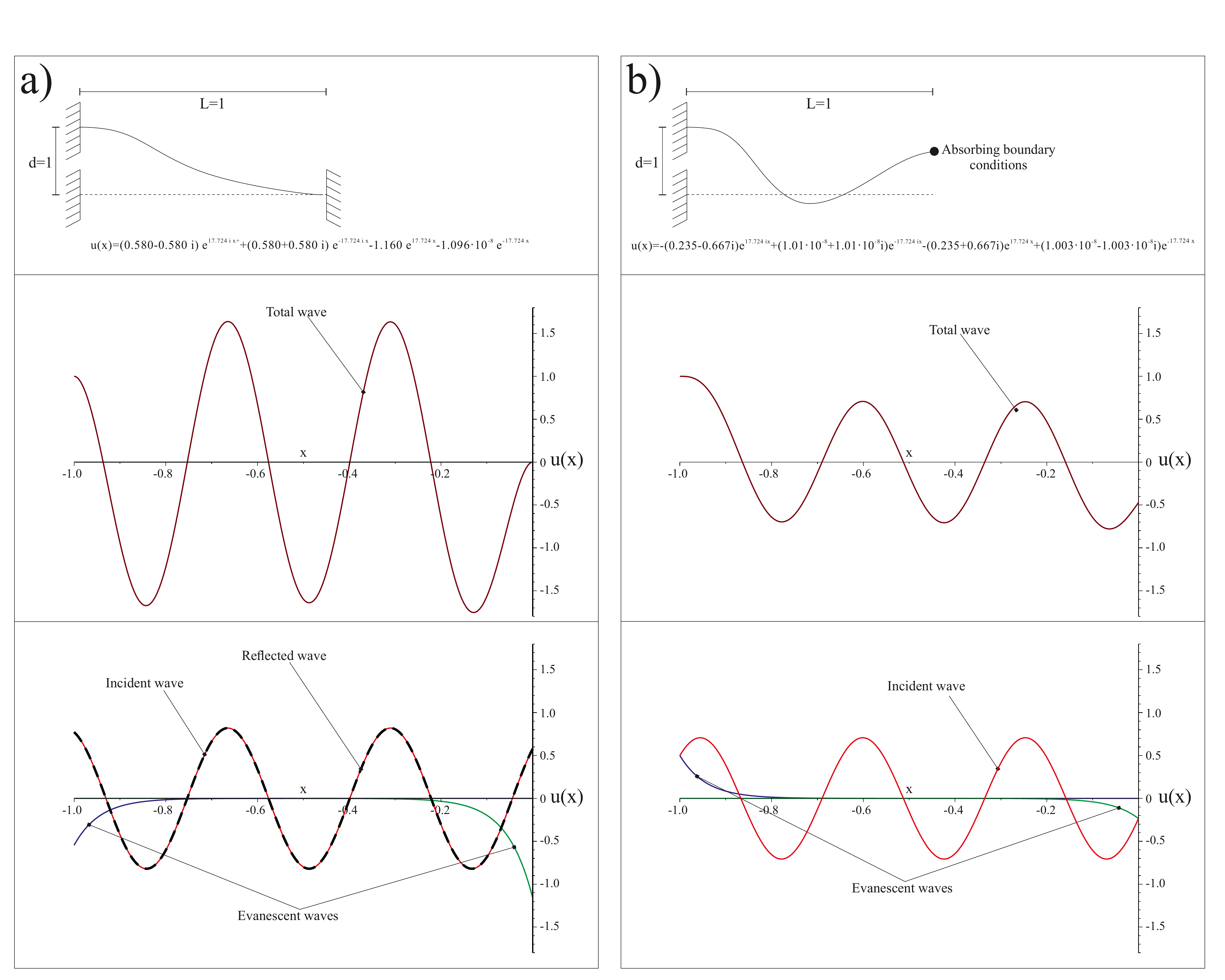}
\caption{\footnotesize a) Euler-Bernoulli beam with clamped boundary conditions on the right end; the incident wave is propagating from left to right (upper part); total wave $u(x)$ (middle part); decomposition of the total wave in incident, reflected and evanescent waves (lower part). b) Euler-Bernoulli beam with absorbing boundary conditions on the right end; the incident wave is propagating from left to right (upper part); total wave $u(x)$ (middle part); decomposition of the total wave in incident  and evanescent waves; note that the reflected wave is absent as a consequence of the absorbing boundary conditions (lower part).}
\label{figappendix}
\end{figure}

Introducing the representation (\ref{eq:gov3}) into the boundary conditions (\ref{eq:gov4}), we get
\begin{equation}
\label{eq:gov5}
C_3^*-C_2^*-1=ia_1(1+C_2^*+C_3^*) \quad \textrm{and} \quad iC_2^*+C_3^*-i = -ia_2(iC_2^*-C_3^*-i).
\end{equation}
The boundary conditions obtained are independent of the angular frequency $\omega$.
It is not possible to set both the coefficients $C_2^*$ and $C_3^*$ equal to zero, so we consider $C_2^*=0$ since it is responsible for the reflecting waves and we evaluated accordingly the parameters $a_1$ and $a_2$ as a function of $C_3^*$:
\begin{equation}
\label{eq:gov6}
a_1=\frac{i-iC_3^*}{C_3^*+1} \quad \textrm{and} \quad a_2=\frac{C_3^*-i}{iC_3^*-1}.
\end{equation}
We obtained for an Euler-Bernoulli beam two approximated absorbing boundary conditions which allow to nullify the reflected waves and to control the evanescent waves by setting the coefficient $C_3^*$. Since the results are independent of the angular frequency $\omega$ they are also valid in the transient regime, in which case the approximated absorbing boundary conditions take the form:
\begin{equation}
\label{eq:gov7}
u''(0,t)=-a_1\dot{u}(0,t) \quad \textrm{and} \quad u'''(0,t)=-a_2\dot{u}'(0,t).
\end{equation}
We consider now the simple case of a Euler-Bernoulli beam, in time harmonic regime, subject to a unit displacement at the left end and clamped at the right end, see Fig.~\ref{figappendix}a, and a second case where the fixed boundary conditions at the right end are replaced with the absorbing boundary conditions obtained previously, see Fig.~\ref{figappendix}b. In the examples we consider $C_3^*=1$, $E=1$, $A=1$, $\rho=1$, $I=1$, $\omega=100\!\cdot\!\pi$ rad/s and span of the beam equal to one. The boundary conditions are $u(-1)=1$, $u'(-1)=0$, $u(0)=0$, $u'(0)=0$ for the clamped case, and $u(-1)=1$, $u'(-1)=0$, $u''(0)=ia_1\beta u(0)$, $u'''(0)=ia_2\beta u'(0)$ for the case with absorbing boundary conditions.

It is possible to see in Fig.~\ref{figappendix}b the reflected wave disappeared and the amplitude of the evanescent wave in proximity of the interface where we applied the approximate absorbing boundary conditions can be tuned changing the value of the coefficient $C_3^*$.

\begin{figure}[!htb]
\centering
\includegraphics[width=160mm]{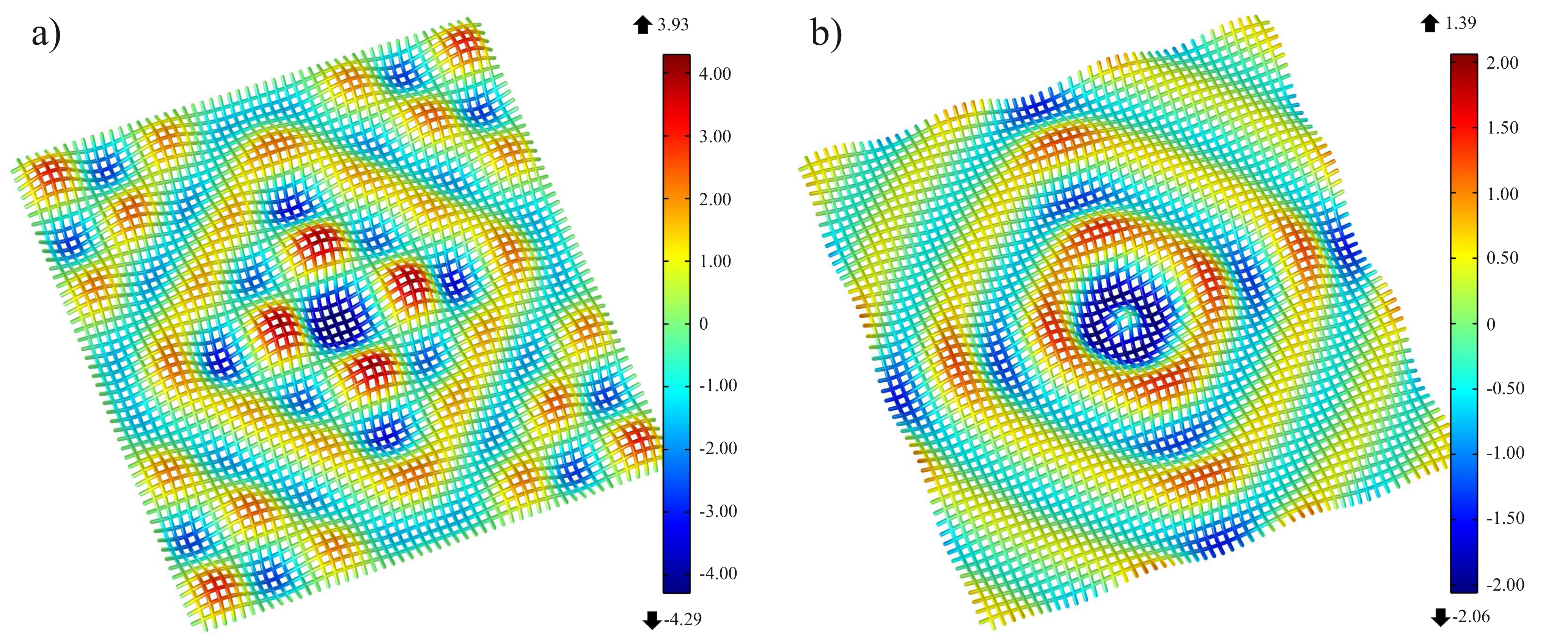}
\caption{\footnotesize Homogeneous Euler-Bernoulli lattice excited by a time-harmonic out-of-plane point force placed at a nodal point in the centre of the network. The angular frequency of the excitation is $\omega=0.02\!\cdot\!\pi$ rad/s. a) Euler-Bernoulli lattice with simple supported boundary conditions on the boundaries. b) Euler-Bernoulli lattice with approximated boundary conditions on the boundaries.}
\label{figappendix1}
\end{figure}

In Fig.~\ref{figappendix1} we show the results of a homogeneous Euler-Bernoulli lattice subject to a out-of-plane point force in the centre, in time harmonic regime for the angular frequency $\omega=0.02\!\cdot\!\pi$ rad/s, and simple supported on the boundaries, Fig.~\ref{figappendix1}a, and the same lattice where we introduced the absorbing boundary conditions, Fig.~\ref{figappendix1}b.

\clearpage

\subsection{Rayleigh beam}

We obtained approximated absorbing boundary condition for the Rayleigh beams as well. The governing equation for time-harmonic flexural waves is:
\begin{equation}
\label{eq:gov8}
EI u''''(x)+\rho I \omega^2 u''(x)-\rho A \omega^2 u = 0.
\end{equation}
The general solution of the equation (\ref{eq:gov8}) is:
\begin{equation}
\label{eq:gov9}
u(x)=C_1e^{i\sqrt{\beta_1}x}+C_2e^{-i\sqrt{\beta_1}x}+C_3e^{\sqrt{\beta_2}x}+C_4e^{-\sqrt{\beta_2}x},
\end{equation}
where $\beta_1=(\sqrt{\rho^2I^2\omega^2+4EI\rho A}+\rho I\omega)\omega/(2 EI)$ and $\beta_2=(\sqrt{\rho^2I^2\omega^2+4EI\rho A}-\rho I\omega)\omega/(2 EI)$. For the Rayleigh beam $C_1e^{i\sqrt{\beta_1}x}$ represents the travelling wave from left to right, $C_2e^{-i\sqrt{\beta_1}x}$ the travelling wave from right to left, $C_3e^{\sqrt{\beta_2}x}$ the evanescent wave decaying at $-\infty$ and $C_4e^{-\sqrt{\beta_2}x}$ the evanescent wave decaying at $+\infty$. For the reflection problem in the semi-infinite beam, in the domain $(-\infty,0)$, the meaning of the coefficients $C_1$, $C_2$, $C_3$ and $C_4$ is the same as for the previous case.

We consider the case where the coefficient responsible for the incident wave is $C_1^*=1$ and the one responsible for the second evanescent wave is $C_4^*=0$, and we determine two boundary conditions in $x=0$ in order to evaluate the two remaining coefficients $C_2^*$ and $C_3^*$:
\begin{equation}
\label{eq:gov10}
u''(0)=a_1 u(0) \quad \textrm{and} \quad u'''(0)=a_2 u'(0).
\end{equation}
It is not possible in this case to have boundary conditions independent of the pulsation $\omega$, thus we determine frequency-dependent parameters $a_1$ and $a_2$, such that the coefficient $C_2^*$ responsible of reflected waves is zero, and the coefficient $C_3^*$ responsible for the evanescent wave is tunable.

Introducing the representation (\ref{eq:gov9}) into the boundary conditions (\ref{eq:gov10}), we get
\begin{equation}
\label{eq:gov11}
-\beta_1+C_3^*\beta_2=a_1(C_3^*+1) \quad \textrm{and} \quad -\beta_1 i\sqrt{\beta_1}+C_3^*\beta_2\sqrt{\beta_2}=a_2(i\sqrt{\beta_1}+C_3^*\sqrt{\beta_2}).
\end{equation}
The boundary conditions obtained are not independent of the angular frequency $\omega$. The values of the parameters $a_1$ and $a_2$ as a function of $C_3^*$ are
\begin{equation}
\label{eq:gov12}
a_1=\frac{-\beta_1+C_3^*\beta_2}{1+C_3^*} \quad \textrm{and} \quad
a_2= \frac{-i\beta_1\sqrt{\beta_1}+C_3^*\beta_2\sqrt{\beta_2}}{i\sqrt{\beta_1}+C_3^*\sqrt{\beta_2}}.
\end{equation}
We obtained for a Rayleigh beam two approximated absorbing boundary conditions, which allow to nullify the reflected waves and to control the evanescent waves by setting the coefficient $C_3^*$.

\begin{figure}[!htb]
\centering
\includegraphics[width=160mm]{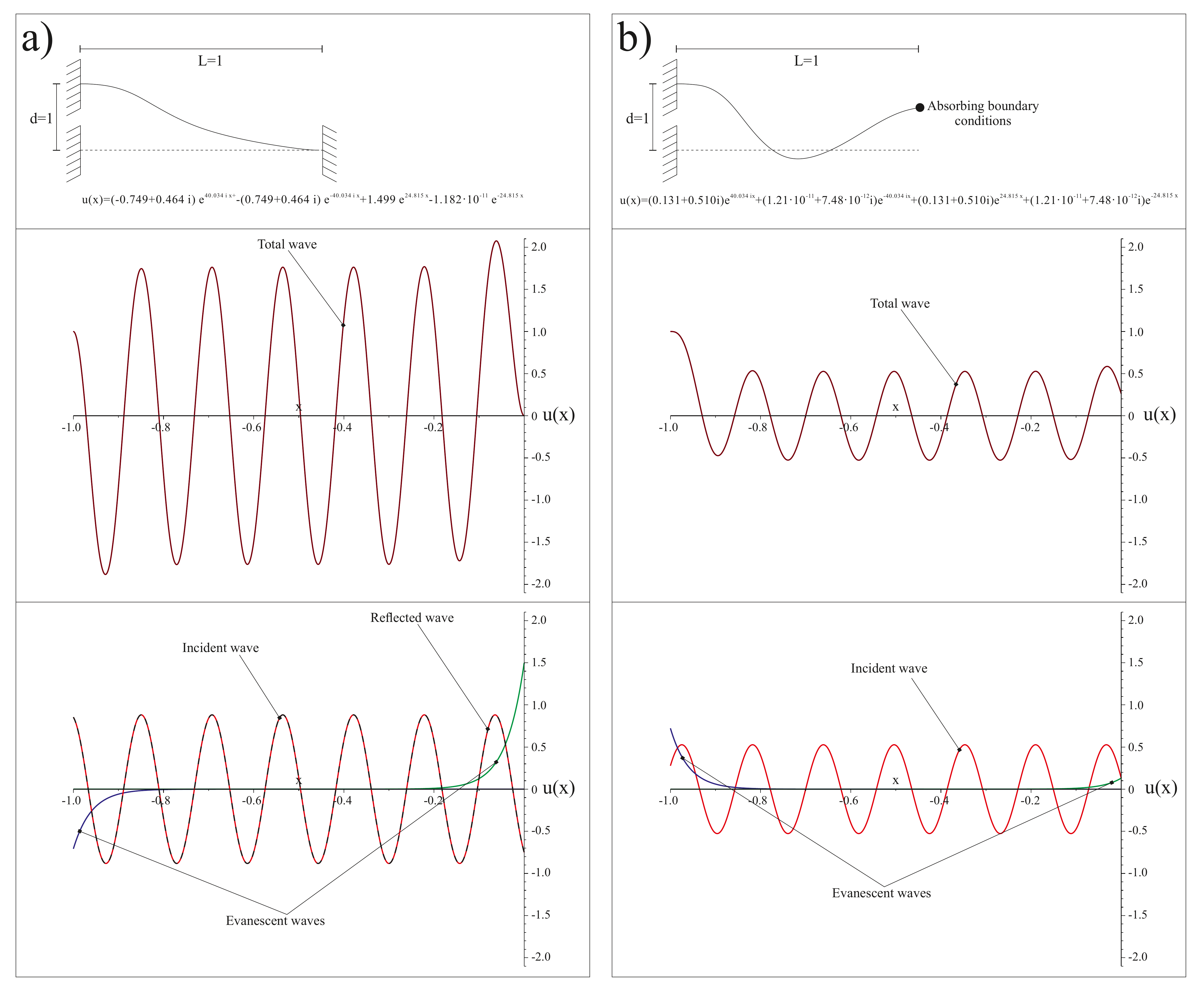}
\caption{\footnotesize a) Rayleigh beam with clamped boundary conditions on the right end; the incident wave is propagating from left to right (upper part); total wave $u(x)$ (middle part); decomposition of the total wave in incident, reflected and evanescent waves (lower part). b) Rayleigh beam with absorbing boundary conditions on the right end; the incident wave is propagating from left to right (upper part); total wave $u(x)$ (middle part); decomposition of the total wave in incident  and evanescent waves; note that the reflected wave is absent as a consequence of the absorbing boundary conditions (lower part)..}
\label{figappendix2-1}
\end{figure}

We consider now the simple case of a Rayleigh beam, in time harmonic regime, subject to a unit displacement at the left end and clamped at the right end, see Fig.~\ref{figappendix2-1}a, and a second case where the fixed boundary conditions at the right end are replaced with the absorbing boundary conditions obtained previously, see Fig.~\ref{figappendix2-1}b. In the examples we consider $C_3^*=1$, $E=1$, $A=1000$, $\rho=1$, $I=1$, $\omega=10\!\cdot\!\pi$ rad/s and span of the beam equal to one. The boundary conditions are $u(-1)=1$, $u'(-1)=0$, $u(0)=0$, $u'(0)=0$ for the clamped case, and $u(-1)=1$, $u'(-1)=0$, $u''(0)=a_1 u(0)$, $u'''(0)=a_2 u'(0)$ for the case with absorbing boundary conditions.

It is possible to see in Fig.~\ref{figappendix2-1}b that the reflected wave disappeared and the amplitude of the evanescent wave in proximity of the interface where we apply the approximate absorbing boundary conditions can be tuned changing the value of the coefficient $C_3^*$.

\begin{figure}[!t]
\centering
\includegraphics[width=160mm]{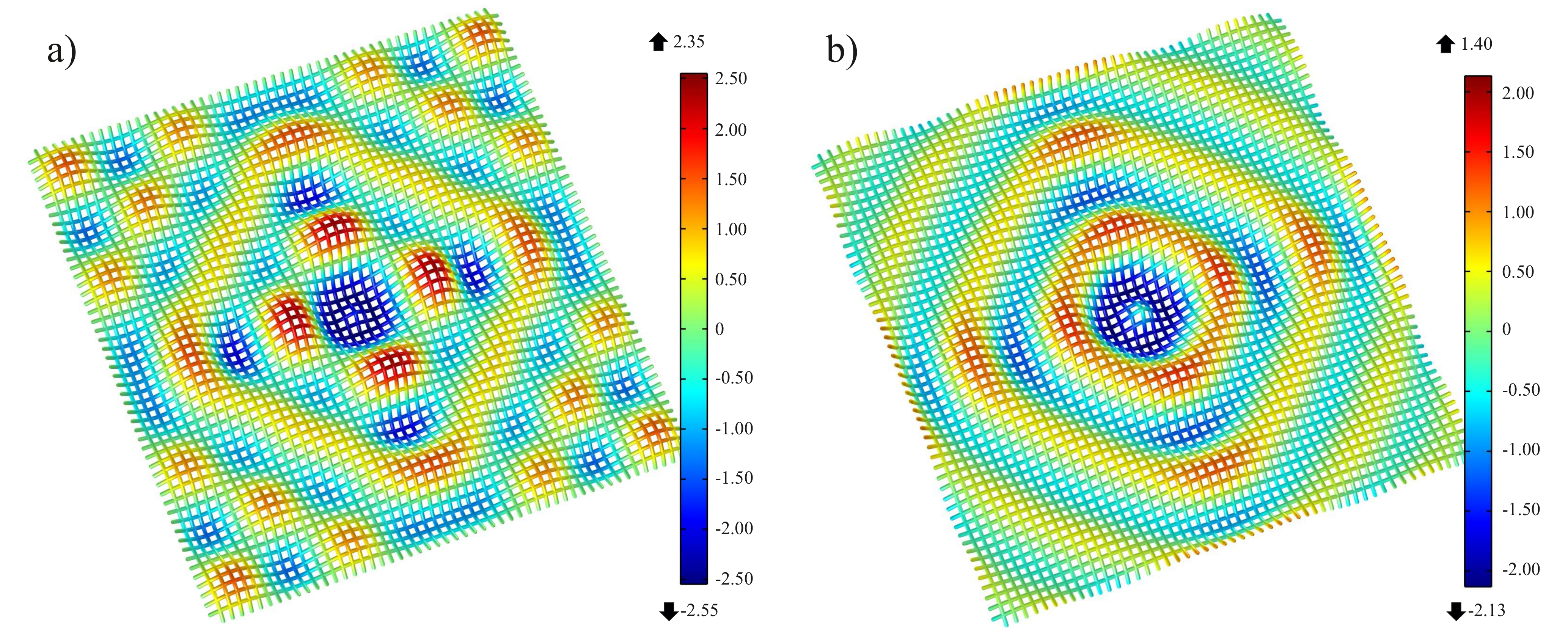}
\caption{\footnotesize Homogeneous Rayleigh lattice excited by a time-harmonic out-of-plane point force placed at a nodal point in the centre of the network. The angular frequency of the excitation is $\omega=0.02\!\cdot\!\pi$ rad/s. a) Rayleigh lattice with simple supported boundary conditions on the boundaries. b) Rayleigh lattice with approximated boundary conditions on the boundaries.}
\label{figappendix3}
\end{figure}

In Fig.~\ref{figappendix3}, we show the results of a homogeneous Rayleigh lattice subject to a out-of-plane point force and simple supported on the boundaries, Fig.~\ref{figappendix3}a, and the same lattice where we introduced the absorbing boundary conditions, Fig.~\ref{figappendix3}b.

By comparing Fig.~\ref{figappendix1}b and Fig.~\ref{figappendix3}b, we notice that the behaviour of the two different lattices is the same after we introduce the absorbing boundary conditions, for the low angular frequency $\omega=0.02\!\cdot\!\pi$ rad/s.

\clearpage

\bibliographystyle{jabbrv_unsrt}
\bibliography
{%
roaz1}

\end{document}

%% file: definitions.tex
\newcommand{\beq}{\begin{equation}}
\newcommand{\eeq}{\end{equation}}

\newcommand{\beqar}{\begin{eqnarray}}
\newcommand{\eeqar}{\end{eqnarray}}
\newcommand{\bit}{\begin{itemize}}
\newcommand{\eit}{\end{itemize}}
\newcommand{\benum}{\begin{enumerate}}
\newcommand{\eenum}{\end{enumerate}}
\newcommand{\barr}{\begin{array}}
\newcommand{\earr}{\end{array}}

\def\XXint#1#2#3{{\setbox0=\hbox{$#1{#2#3}{\int}$}
   \vcenter{\hbox{$#2#3$}}\kern-.5\wd0}}

\def\b0{\mbox{\boldmath $0$}}

\def\f0{\ensuremath{\mathbb{O}}}

